\documentclass[twocolumn,nofootinbib]{revtex4-1}
\usepackage{graphicx}
\usepackage{epstopdf}
\usepackage{amsmath}

\newcommand{\pr}{p_R}
\newcommand{\cum}[2]{\left \langle {#1}^{#2} \right \rangle_{\!c}}
\newcommand{\delN}{\Delta N}
\newcommand{\netp}{{p\!-\!\bar p}}
\newcommand{\tfo}{T_{\mathrm{fo}}}
\newcommand{\dif}{\mathrm{d}}
\newcommand{\chf}{\varphi(i\xi)}
\newcommand{\mom}[1]{{\mu_{#1}^\prime}}
\newcommand{\fm}[1]{{F_{#1}}}
\newcommand{\nrfm}[1]{\left \langle \frac{N_R!}{(N_R - #1)!} \right \rangle}
\newcommand{\eix}{e^{i\xi}}
\newcommand{\Md}[1]{M^{(#1)}}
\usepackage[shortlabels]{enumitem}
\begin{document}
\title{Proton number fluctuations in partial chemical equilibrium}
\author{Boris Tomasik$^{a,b}$}
%\author{Boris Tom\'a\v{s}ik$^{a,b}$}
\author{Paula Hillmann$^{c,d,e,f}$}
\author{Marcus Bleicher$^{c,d,e,f}$}
\affiliation{$^a$\v{C}esk\'e vysok\'e u\v{c}en\'i technick\'e v Praze, 
FJFI, B\v{r}ehov\'a 7, 11519 Praha 1, Czechia}
\affiliation{$^b$Univerzita Mateja Bela, Tajovsk\'eho 40, 97401 Bansk\'a Bystrica, Slovakia}
%\affiliation{$^c$Frankfurt Institute for Advanced Studies, 
%Johann Wolfgang Goethe-Universit\"at, Ruth-Moufang-Strasse 1, 60438 Frankfurt am Main, Germany}
\affiliation{$^c$Institut f\"ur Theoretische Physik, 
Johann Wolfgang Goethe-Universit\"at, Max-von-Laue-Strasse 1, 60438 Frankfurt am Main, Germany}
\affiliation{$^d$Helmholtz Research Academy Hesse for FAIR (HFHF), Campus Frankfurt,Max-von-Laue-Str.  12, 60438 Frankfurt am Main, Germany}
\affiliation{$^e$GSI Helmholtz Center, Planckstr.~1, 64291 Darmstadt, Germany}
\affiliation{$^f$ John von Neumann-Institut f\"ur Computing, Forschungszentrum J\"ulich,
52425 J\"ulich, Germany}


\begin{abstract}
We calculate volume-independent ratios of cumulants of the net-proton number distribution up to  sixth  order  in  a  fireball  that  cools  down  after  the  chemical  freeze-out.   A  hadron  resonance gas  model  is  used  together  with  the  assumption  of  partial  chemical  equilibrium,  which  fixes  the number  of  observed  stable  hadrons  after  the  chemical  freeze-out.   It  is  shown  that  due  to  only weak departure from the statistical Boltzmann distribution, also the volume-independent ratios of higher-order cumulants of the net-proton number show only weak dependence on the temperature.This observation supports the possibility to measure non-critical cumulants at chemical freeze-out even after subsequent cooling in the hadronic phase.  Cumulants of the net-baryon number behave similarly, while those for the kaon number vary more strongly with the temperature.  Our results are relevant for the current fluctuation studies of the RHIC-BES runs.
\end{abstract}
\maketitle

%%%%%%%%%%%%%%%%%%%%%%%%%%%%%%%%%%%%%%%%%%%%%%%%

\section{Introduction}

Mapping the structure of the phase diagram of strongly interacting matter is a major objective of today's high-energy nuclear physics. 
According to the current understanding of Quantum Chromo Dynamics (QCD) one expect that at small net-baryon density and high enough temperature hadronic matter undergoes a rapid but smooth crossover to a deconfined quark-gluon plasma
\cite{Aoki:2006we}. One anticipates that this crossover ends in a critical end point (second order phase transition) and continues to even higher baryon densities in a first order transition line. Currently, a tremendous amount of theoretical and experimental activity is focused on the search for the exact position of this critical 
point (\cite{Bzdak:2019pkr,Bluhm:2020mpc} and references therein). 

Among the most promising observables in this context are the fluctuations of the baryon number which can be accessed for equilibrated matter within the grand canonical ensemble, e.g. on the lattice or in a statistical model \cite{Bazavov:2020bjn,Nahrgang:2014fza}. 
The fluctuations are customarily quantified by the cumulants of the net-baryon number distribution. 
It has been shown that the baryon number fluctuations scale with high powers of the 
correlation length, this makes higher-order cumulants particularly sensitive to the vicinity of the critical point, in which the correlation 
length diverges  \cite{Hatta:2002sj}.  

Unfortunately, connecting those theoretical findings to real experimentally accessible observables is far from straightforward:
\begin{itemize}
    \item detector efficiencies and acceptance cuts might play a role \cite{He:2017zpg,Sombun:2017bxi},
    \item cluster production may influence the results \cite{Feckova:2015qza},
    \item protons are used as a proxy for baryon number~\cite{Hatta:2003wn,Kitazawa:2012at},
    \item the duration and volume of the system might be too small to reach grand canonical equilibrium \cite{Bebie:1991ij,Hirano:2002ds,Rapp:2002fc},
\end{itemize}
to just name a few of the challenges.

Very interesting results were recently
published by the STAR collaboration \cite{Abdallah:2021fzj,Abdallah:2021zhr}, which indicate a dramatic rise of the fourth-order cumulant as the energy of Au+Au 
collisions has been lowered to 7.7~GeV per nucleon pair in framework 
of the RHIC Beam Energy Scan program. 
No theoretical 
explanations that can explain this observation exist to this date. 

In parallel to identifying the reason of this observation, it is crucial to improve the calculations of the expected behaviour, namely the baseline of
non-critical scenarios 
\cite{Bzdak:2012an,Nahrgang:2014fza,Feckova:2015qza,Bzdak:2016sxg,Braun-Munzinger:2016yjz,Bzdak:2018uhv,Braun-Munzinger:2020jbk}. 
Only in this way the strength of the signal above the non-critical background can be reliably quantified. 
A natural choice of a baseline is the statistical model of a hadron resonance gas \cite{Nahrgang:2014fza,Dashen:1969ep,Cleymans:1998fq,Andronic:2017pug}. It has been frequently used in the description 
of the hadron abundances produced in nuclear collisions at various energies where it provides a reasonable fit to measured data. 
Technically, one deals with first-order cumulants in this kind of analysis. 

The statistical hadron resonance gas model combined with lattice QCD simulations has also been used in fitting higher-order cumulants measured by STAR
in order to extract the freeze-out temperature from these data \cite{Alba:2014eba,Alba:2015iva}. 
While the overall agreement is good, a slight disagreement with the thermal parameters obtained 
from fitting the abundances (first order cumulant) has been reported. In \cite{Sochorova:2018ojd} it has been pointed out, that this may be a consequence
of fast expansion and cooling which drives the particle number distribution out of equilibrium so that its different cumulants cannot 
be assigned to a unique value of the temperature. 
 
On the other hand, since the fireball stays connected after the chemical freeze-out and cools further down, Partial Chemical Equilibrium
(PCE) \cite{Bebie:1991ij,Hirano:2002ds,Rapp:2002fc} is often assumed. It keeps the (effective) numbers of long living hadrons 
%(hadrons up to $\omega$ are considered as stable) 
fixed at the values they had at 
the chemical freeze-out. PCE requires fast equilibration between the stable hadron sort and all resonances that decay into this 
sort of hadrons. 

%As pointed out above, 
Baryon number conservation and the fact that only some of the baryons are detected has an important 
impact on the observed result. Recently, it has been shown how these effects are built into
the statistical models \cite{Braun-Munzinger:2020jbk,Vovchenko:2020tsr,Vovchenko:2020gne}.

In this paper we 
%answer this question and 
explore how the cumulants behave in a cooling fireball that obeys the Partial Chemical Equilibrium (PCE) scenario. The paper is structured in the following way: First, we introduce the PCE scenario in the next Section, then we explain how the net-proton number cumulants are calculated
from a generating function and the grand-canonical partition function. 
We show results for the grand canonical PCE model applied to central Au+Au collisions from 
the RHIC Beam Energy Scan program between $\sqrt{s_{NN}} = 7.7$ GeV and 200 GeV for the net-proton number, 
net-baryon number, and also the number $(K^+-K^-)$ in Section~\ref{s:results}.
Their implications 
are summarised in the concluding Section~\ref{s:conc}. We also include two appendices where we display in detail the way how the cumulants 
are derived and calculated. 

%%%%%%%%%%%%%%%%%%%%%%%%%%%%%%%%%%%%%%%%%%%%%%%%

\section{Formalism and the model}
\label{s:model}

%%%%%%%%%%%%%%%%%%%%
\subsection{Cumulants and moments of the number distribution}
\label{ss:cumulants}

Generally, cumulants characterise a  probability distribution. They are calculated from the cumulant-generating function, defined as
\begin{equation}
K(i\xi) = \ln \sum_{N=0}^{\infty} e^{i\xi N} P(N)\,  , 
\label{e:cumder}
\end{equation}
where $P(N)$ is the (discrete) probability distribution and $\xi$ is an auxiliary parameter. Cumulants are obtained by taking 
derivatives
\begin{equation}
\cum{(\delN)}{l} = \left . \frac{\mathrm{d}^l K(i\xi)}{\mathrm{d} (i\xi)^l}\right |_{\xi=0}\,  .
\label{e:cumdef}
\end{equation}
We will use the notation $\cum{(\delN)}{l}$ to denote the cumulant of $l$-th order, for $l>1$. The case $l=1$ is usually denoted differently 
and is called the mean  of the distribution
\begin{equation}
M = \langle N\rangle = \cum{N}{} = \left . \frac{\mathrm{d} K(i\xi)}{\mathrm{d} (i\xi)} \right |_{\xi=0}\, .
\end{equation}
The first three cumulants are identical with the central moments, while the higher moments differ from them. 

Ultimately, we will be interested in the \emph{net} proton number, which is the difference of two random variables: 
number of protons and number of antiprotons. 
For a difference of independent random variables, the 
cumulants  follow the relation
\begin{equation}
\cum{(\delN_\netp)}{l} = \cum{(\delN_p)}{l} + (-1)^l \cum{(\delN_{\bar p})}{l}\,  .
\label{e:cumppbar}
\end{equation}

It follows from Eq.~(\ref{e:cumdef}) that cumulants appear naturally in statistical physics. In the grand-canonical formalism, 
cumulants of the number distribution of a given sort of particles $j$ are calculated from the derivatives of the logarithm of the 
partition function $\cal Z$ with respect to the corresponding chemical potential $\mu_j$ scaled by the temperature
\begin{equation}
\cum{(\delN_j)}{l} = \frac{\partial^l \ln {\cal Z}(V,T,m_j,\mu_j)}{\partial (\mu_j/T)^l}\,  ,
\label{e:cumpartitionsum}
\end{equation}
In the grand canonical ensemble these cumulants are solely due to fluctuations that are caused by the exchange of quantum numbers with the heatbath. 
Hence, it  describes fluctuations of the baryon number or strangeness, which are conserved in any microscopic process within 
the system. However, Eq.~(\ref{e:cumpartitionsum}) does \emph{not} apply to proton number fluctuations, because proton number fluctuations can also be caused by the stochastic decay of a resonance into protons. 

{\em Susceptibilities} of a conserved quantum number are introduced by scaling out the volume and making them dimensionless
\begin{equation}
\chi_l^{(j)} = \frac{1}{VT^3} \frac{\partial^l \ln {\cal Z}(V,T,m_j,\mu_j)}{\partial (\mu_j/T)^l}\, .
\end{equation}
This makes them convenient for theoretical calculations. 

The cumulants of the number distribution can be experimentally accessed via the event-by-event moments of the particle number distribution. Customarily, standardised moments are used for higher orders:
the variance, the skewness, and the kurtosis $\left( \Delta N=N_i-\left< N\right> \right)$:
\begin{eqnarray}
\sigma^2 & = & \cum{(\delN)}{2}\,  ,\\
S & = & \frac{\cum{(\delN)}{3}}{{\cum{(\delN)}{2}}^{3/2}}\,  , \\
\kappa & = & \frac{\cum{(\delN)}{4}}{{\cum{(\delN)}{2}}^{2}}\,  .
\end{eqnarray}
The next two orders are called hyper-skewness and hyper-kurtosis
\begin{eqnarray}
S ^H& = & \frac{\cum{(\delN)}{5}}{{\cum{(\delN)}{2}}^{5/2}}\,  , \\
\kappa^H & = & \frac{\cum{(\delN)}{6}}{{\cum{(\delN)}{2}}^{3}}\,  .
\end{eqnarray}
Such moments still depend on the volume. 
In order to streamline the comparison of measured moments and calculated susceptibilities, 
volume-independent ratios are particularly suitable
\begin{align}
\nonumber
\frac{\chi_2}{\chi_1} & = \frac{\cum{(\delN)}{2}}{\cum{N}{}} =   \frac{\sigma^2}{M} &
\frac{\chi_3}{\chi_2} & = \frac{\cum{(\delN)}{3}}{\cum{(\delN)}{2}} = S\sigma \\
\nonumber
\frac{\chi_4}{\chi_2} & = \frac{\cum{(\delN)}{4}}{\cum{(\delN)}{2}} = \kappa\sigma^2 &
\frac{\chi_5}{\chi_1} & = \frac{\cum{(\delN)}{5}}{\cum{N}{}} = \frac{S^H\sigma^5}{M} \\
\frac{\chi_5}{\chi_2} & = \frac{\cum{(\delN)}{5}}{\cum{(\delN)}{2}} = S^H\sigma^3 &
\frac{\chi_6}{\chi_2} & = \frac{\cum{(\delN)}{6}}{\cum{N}{}} =\kappa^H\sigma^4\,  .
\label{e:volind}
\end{align}
Note that we listed two different combinations of $\chi_5$, since both appear in the literature. 

%%%%%%%%%%%%%%%%%%%%%%%%%%%%%%
\subsection{Cumulants of the proton number distribution}
\label{ss:pcums}

In contrast to the baryon number, the proton number is not a conserved quantum number. Thus, the cumulants of its distribution cannot be calculated by just taking derivatives of the partition function. In addition to direct proton production, protons can also originate from decays of resonances, which 
is a random process that contributes to the proton number fluctuations. 

We demonstrate in the Appendix that the (anti-)proton number cumulants can be calculated from this cumulant-generating function:
\begin{equation}
K(i\xi) = \sum_R \ln \left \{ \sum_{N_R=0}^\infty P_R(N_R) \left ( e^{i\xi}\pr + (1-\pr) \right )^{N_R} \right \}\,  .
\end{equation}
Here, the first sum counts all sorts of resonances that contribute to proton production and the second sum runs through
the numbers of resonances of a given sort. The probability to have $N_R$ resonances is $P_R(N_R)$, and 
$\pr$ ($0\le \pr \le 1$) denotes the mean number of protons produced in decays of resonance $R$. If a resonance decays via a chain 
of subsequent decays, then $\pr$ counts the average number after all decays have happened. Cumulants 
are calculated according to Eq.~(\ref{e:cumder}). From this, we derive the first six cumulants of the proton number
(see Appendix for details):
\begin{widetext}
\begin{subequations}
\label{e:allpcums}
\begin{eqnarray}
\cum{N_p}{} & = & \sum_R \pr \cum{N_R}{}\,  \\
\cum{(\delN_p)}{2} & = & \sum_R \left [ \pr^2 \cum{(\delN_R)}{2} + \pr(1-\pr) \cum{N_R}{} \right ] \, , \\
\cum{(\delN_p)}{3} & = & \sum_R \left [  \pr^3 \cum{(\delN_R)}{3}  + 3 \pr^2(1-\pr)  \cum{(\delN_R)}{2} 
+ \pr(1-\pr)(1-2\pr)\cum{N_R}{} \right ] \,  , \\
\cum{(\delN_p)}{4} & = & \sum_R \Bigl [
\pr^4\cum{(\delN_R)}{4} + 6\pr^3(1-\pr)\cum{(\delN_R)}{3}
 + \pr^2(1-\pr)(7-11\pr)\cum{(\delN_R)}{2} 
 \nonumber \\
 && \qquad \qquad  {}
+ \pr(1-\pr)(1-6\pr+6\pr^2) \cum{N_R}{} \Bigr ] \,  , \\
\cum{(\delN_p)}{5} & = & \sum_R \Bigl [
\pr^5\cum{(\delN_R)}{5} +10 \pr^4(1 - \pr)  \cum{(\delN_R)}{4} 
+ 5 \pr^3 (1-\pr)( 5-7\pr) \cum{(\delN_R)}{3}
\nonumber \\ 
&& \qquad {}
+5 \pr^2 (1-\pr)(10\pr^2 - 12\pr+3) \cum{(\delN_R)}{2} 
\nonumber \\ 
&& \qquad {}
+  \pr(1 - \pr)  (1 - 2 \pr)  (12\pr^2 - 12 \pr  +1)  \cum{N_R}{} \Bigr ]\, ,
\\
\cum{(\delN_p)}{6} & = & \sum_R \Bigl [
\pr^6\cum{(\delN_R)}{6} +15\pr^5(1-\pr)\cum{(\delN_R)}{5}+5\pr^4(1-\pr)(13-17\pr)\cum{(\delN_R)}{4}
\nonumber \\ 
&& \qquad {}
+ 15\pr^3(1-\pr)(15\pr^2-20\pr+6)\cum{(\delN_R)}{3} 
\nonumber \\ 
&& \qquad {}
- \pr^2(1-\pr)(  274 \pr^3 -476 \pr^2  +239 \pr - 31  )\cum{(\delN_R)}{2}
\nonumber \\ 
&& \qquad {}
+ \pr(1-\pr)( 120 \pr^4 - 240 \pr^3 + 150 \pr^2  - 30 \pr + 1)\cum{N_R}{}  \Bigr ]\,  ,
\end{eqnarray}
\end{subequations}
\end{widetext}
where the sums go through all resonance species and $\cum{(\delN_R)}{l}$ is the $l$-th cumulant of the 
number distribution of the resonance $R$
\begin{equation}
\label{e:Rcum}
\cum{(\delN_R)}{l} =\frac{\partial^l \ln {\cal Z}_R}{\partial (\mu_R/T)^l}\,  .
\end{equation}
Here, ${\cal Z}_R$ is the partition function for the resonance species $R$, and $\mu_R$ is the chemical potential 
of this resonance species.
These expressions agree with \cite{Nahrgang:2014fza}, where 
they were calculated up to fourth order. 

Note that the sums in Eqs.~(\ref{e:allpcums}) also include direct thermal production of protons, $R=p$. In that case
$\pr = 1$ and $N_R$ becomes simply the direct proton number.

From the relations (\ref{e:allpcums}) one can construct all volume-independent ratios that are being measured. 

In the next section we will evaluate these expressions according to the grand canonical hadron resonance gas model under the assumption 
of partial chemical equilibrium.

%%%%%%%%%%%%%%%%%%%%%%%%%%%
\subsection{Hadron resonance gas model}
\label{ss:HRG}

We shall work with the hadron resonance gas model, i.e. the interaction of the hadrons are accounted for by the inclusion of resonances
into the partition function \cite{Dashen:1969ep}.
In this approximation, both stable hadrons and resonances are assumed to have a small width compared to the temperature, i.e. $\Gamma ^{\mathrm{tot}}/T\ll 1$. For the present calculation we assume a vanishing $\Gamma ^{\mathrm{tot}}$. The logarithm of the partition function is given as
\begin{eqnarray}
\ln {\cal Z} & = & \sum_R \ln {\cal Z}_R(T,V,m_R,\mu_R)
\nonumber \\ 
& = &
\sum_R (\pm 1) \frac{g_R V}{2\pi^2} 
\\
\nonumber && \qquad \times
 \int_0^\infty dk\, k^2\, \ln \left ( 1 \pm e^{\mu_R/T} e^{-\sqrt{m_R+k^2}/T} \right )\, ,
\end{eqnarray}
where the sum runs over hadronic species including resonance states. The upper and lower signs are for fermionic and bosonic species, 
respectively. Furthermore, $g_R$ is the spin degeneracy (isospin is treated explicitly), $m_R$ the mass, and we include
here $\mu_R$ as the chemical potential for each resonance species separately. The integral is conveniently expressed with the help of 
an infinite sum
\begin{equation}
\ln {\cal Z} = \sum_R \frac{g_RV}{2\pi^2} m_R^2 T \sum_{j=1}^\infty \frac{(\mp 1)^{j-1}}{j^2} e^{j\mu_R/T} K_2\left ( \frac{jm_R}{T} \right )\, ,
\end{equation}
where $K_2$ is the modified Bessel function of second kind and order 2. For large arguments $K_2$ can be approximated by a (decreasing) exponential, which allows to limit the sum to the first terms in numerical calculations.  

The cumulants can now be  directly calculated using Eq.~(\ref{e:Rcum}) as
\begin{eqnarray}
\cum{N_R}{} & = & \frac{g_RV}{2\pi^2} m_R^2 T 
\nonumber \\ &&  \times 
\sum_{j=1}^\infty \frac{(\mp 1)^{j-1}}{j} e^{j\mu_R/T} K_2\left ( \frac{jm_R}{T} \right )\, , 
\label{e:meanR}
\\
\cum{(\delN_R)}{l} & = & \frac{g_RV}{2\pi^2} m_R^2 T 
\nonumber \\ &&  \times
\sum_{j=1}^\infty (\mp 1)^{j-1}{j}^{l-2} e^{j\mu_R/T} K_2\left ( \frac{jm_R}{T} \right )\, .
\label{e:cumR}
\end{eqnarray}
%%%%%%%%%%%%%%%%%%%%%%%%%%%

\subsection{Partial chemical equilibrium}
\label{ss:PCE}
A heavy ion reaction typically proceeds via 3 stages: the initial stage, the compression stage and the freeze-out stage. The freeze-out can be split into two distinct phases, the chemical freeze-out where inelastic flavour changing processes end, and the kinetic freeze-out where all interactions cease.

The multiplicities of identified stable hadrons observed in ultrarelativistic nuclear collisions indicate that they have been 
fixed during chemical freeze-out at a temperature close to the transition from confined to deconfined matter. 
The transverse momentum distributions indicate that the kinetic freeze-out happens at a substantially lower temperature \cite{Melo:2019mpn}. 
Hence, the expansion and cooling between chemical and thermal freeze-out must happen in such a way that 
the effective average number of each stable species is conserved \cite{Bebie:1991ij}. The term "effective" number indicates the inclusion of those hadrons 
which would be produced when all unstable resonances decay, hence for hadron species $h$
\begin{equation}
\cum{N_h^{\mathrm{eff}}}{} = \sum_R p_{R\to h} \cum{N_R}{}\,  .
\end{equation}
Here, the sum runs over all hadron and resonance species, and $p_{R\to h}$ is the average number of hadrons $h$ 
resulting from a decay of resonance $R$. Note that this sum also includes directly produced $h$'s ($R=h$), for 
which formally $p_{R\to h} = 1$. For protons, $p_{R\to h} = \pr$.

In partial chemical equilibrium, resonances remain equilibrated with their decay daughter particles. Thus, their chemical potentials are equal to the sums of the chemical potentials of their daughter particles. For example, since $\Delta^{++}\to p+\pi^+$, we have 
\[
\mu_{\Delta^{++}} = \mu_p + \mu_{\pi^+}\,  .
\]
For resonances with multiple decay channels, the daughter chemical potentials are multiplied with the mean number 
of stable hadrons $h$ resulting from a decay of an $R$
\begin{equation}
\mu_R = \sum_h p_{R\to h} \mu_h\,  .
\end{equation}
The sum runs through all stable hadrons (which are among the daughters of $R$). 

To conserve $\cum{N_h^{\mathrm{eff}}}{}$, each stable species obtains its own $\mu_h$ as the system cools down. 
To formulate   the calculation of $\mu_h(T)$, conservation of $\cum{N_h^{\mathrm{eff}}}{}$ itself is not sufficient, because one 
does not know the volume, which also enters into the multiplicity.
This problem can be solved by assuming isentropic expansion \cite{Bebie:1991ij}. The entropy, calculated as
\begin{equation}
\label{e:S}
S = \sum_R \frac{V P_R + E_R - \cum{N_R}{} \mu_R}{T}\,  ,
\end{equation}
is an extensive quantity, as well, and hence the ratio $\cum{N_h^{\mathrm{eff}}}{}/S$ does not depend on the volume. 
In Eq.~(\ref{e:S}), the sum includes all  hadrons and resonances, and the partial pressure and energy can be 
calculated as 
\begin{eqnarray}
P_R & = &\frac{T \ln {\cal Z}_R}{V} ,
\\
E_R& = & \frac{g_R V}{2\pi^2} \int_0^\infty dk\, k^2\, \sqrt{k^2+m_R^2} \nonumber 
\\ && \quad \quad 
\times \left ( \exp\left (\frac{\sqrt{k^2 + m_R^2} - \mu_R}{T}\right ) \pm 1 \right )^{-1}\,   .
\end{eqnarray}
Hence, $\mu_h(T)$ can be determined from the condition
\begin{equation}
\frac{\cum{N_h^{\mathrm{eff}}(T)}{}}{S(T)} = \left . \frac{\cum{N_h^{\mathrm{eff}}(T)}{}}{S(T)} \right |_{T = \tfo}\,  ,
\end{equation}
where $\tfo$ is the chemical freeze-out temperature. The chemical potential is evolved from the chemical freeze-out 
temperature down to lower temperatures towards the kinetic freeze-out. 

Note that in this model, baryons and their antibaryons are considered separately, so that no annihilation is assumed. 

We evolve the values of the chemical potentials starting from chemical freeze-out as extracted from the multiplicity 
ratios by STAR collaboration in \cite{Adamczyk:2017iwn}. In that work, the grand-canonical ensemble with strangeness 
undersaturation has been used, so the partition function was
\begin{multline}
\ln {\cal Z}  = 
\sum_i (\pm 1) \frac{g_r V}{2\pi^2} 
\\
 \times
 \int_0^\infty dk\, k^2\, \ln \left ( 1 \pm \gamma_s^{|S_i|}e^{\mu_i/T} e^{-\sqrt{m_i+k^2}/T} \right )\, ,
\end{multline}
where the sum goes over all species $i$. The chemical potential 
\begin{equation}
\mu_i = B_i \mu_B + S_i \mu_S\,  ,
\end{equation}
where $B_i$ and $S_i$ are the baryon number and the strangeness of species $i$ and $\mu_B$, $\mu_S$ are
the corresponding chemical potentials. The chemical potential due to isospin has been neglected. Strangeness undersaturation 
is expressed by the parameter $\gamma_s$. The parameters for different energies \cite{Adamczyk:2017iwn} are listed in 
Table~\ref{t:foparams}.
%
%%%%%%%%%%%%%%%%%%%%%%%%%%%%%%
\begin{table}
\caption{Chemical freeze-out parameters for different collision energies \cite{Adamczyk:2017iwn,Abdallah:2021fzj}}
\label{t:foparams}
\begin{center}
\begin{tabular}{|c|c|c|c|c|}
\hline
$\sqrt{s_{NN}}$ [GeV] & $\tfo$ [MeV] & $\mu_B$ [MeV]  & $\mu_S$ [MeV] & $\gamma_s$ \\
\hline
\hline
7.7	&	144.3	&	398.2	&	89.5	&	0.95\\
11.5	&	149.4	&	287.3	&	64.5	&	0.92\\
14.5	&	151.6	&	264.0	&	58.1	&	0.94\\
19.6	&	153.9	&	187.9	&	43.2	&	0.96\\
27.0	&	155.0	&	144.4	&	33.5	&	0.98\\
39.0	&	156.4	&	103.2	&	24.5	&	0.94\\
54.4	&	160.0	&	83.0		&	18.7	&	0.94\\
62.4	&	160.3	&	69.8		&	16.7	&	0.86\\
200	&	164.3	&	28.4		&	5.6	&	0.93\\
\hline
\end{tabular} 
\end{center}
\end{table}
%%%%%%%%%%%%%%%%%%%%%%%%%%%%%%
%
For $\sqrt{s_{NN}} = 14.5$~GeV and 54.4~GeV the values were not included in \cite{Adamczyk:2017iwn}.
Therefore, in these cases we have taken the temperatures and the baryochemical potentials from \cite{Abdallah:2021fzj}.
For $\gamma_s$ we have assumed the value given by weighted average of the values at other energies and $\mu_S$ 
has been determined from the requirement of strangeness neutrality. 

\begin{figure}[t!]
\begin{center}
	\includegraphics[width=0.48\textwidth]{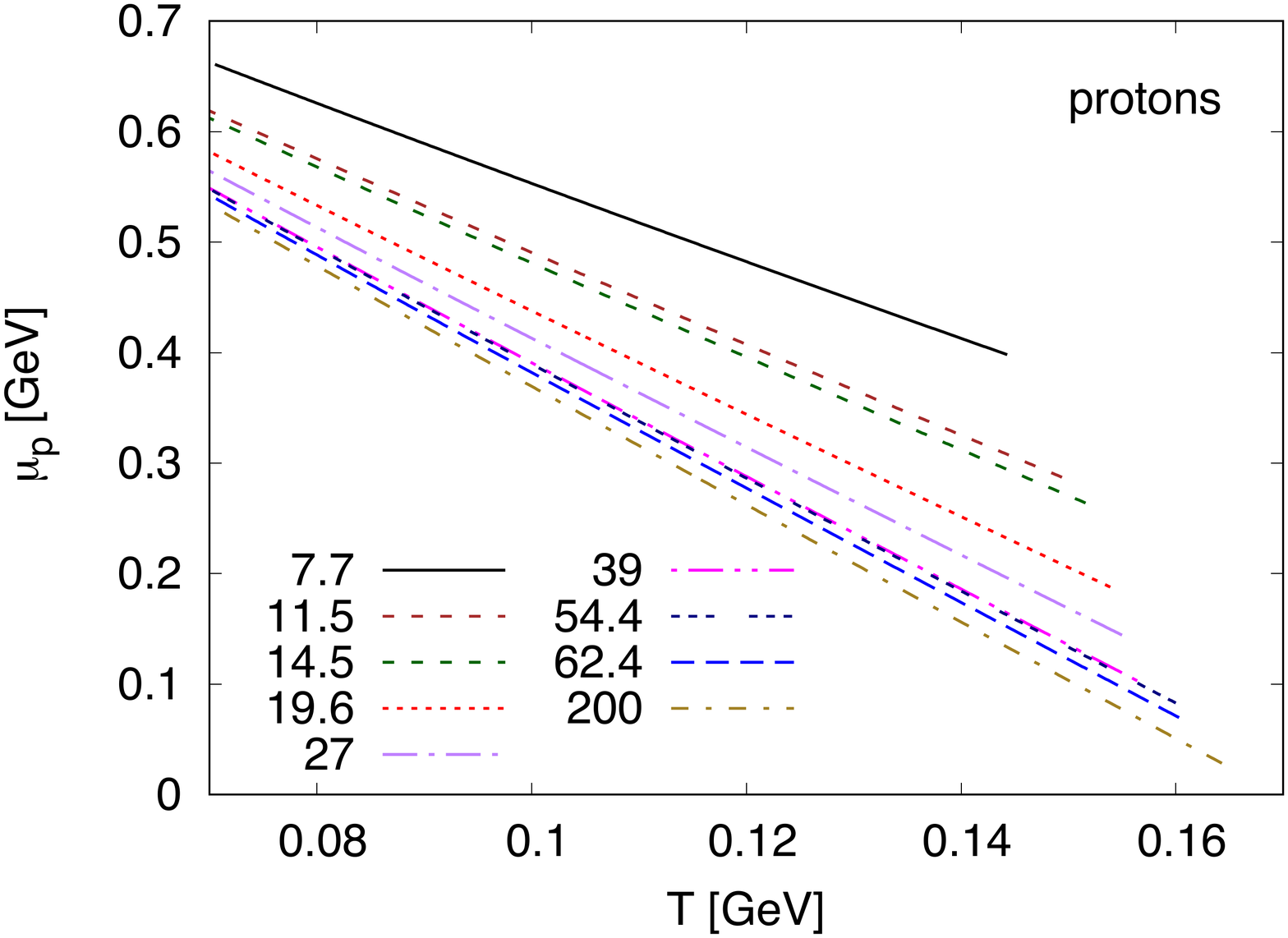}
	\includegraphics[width=0.48\textwidth]{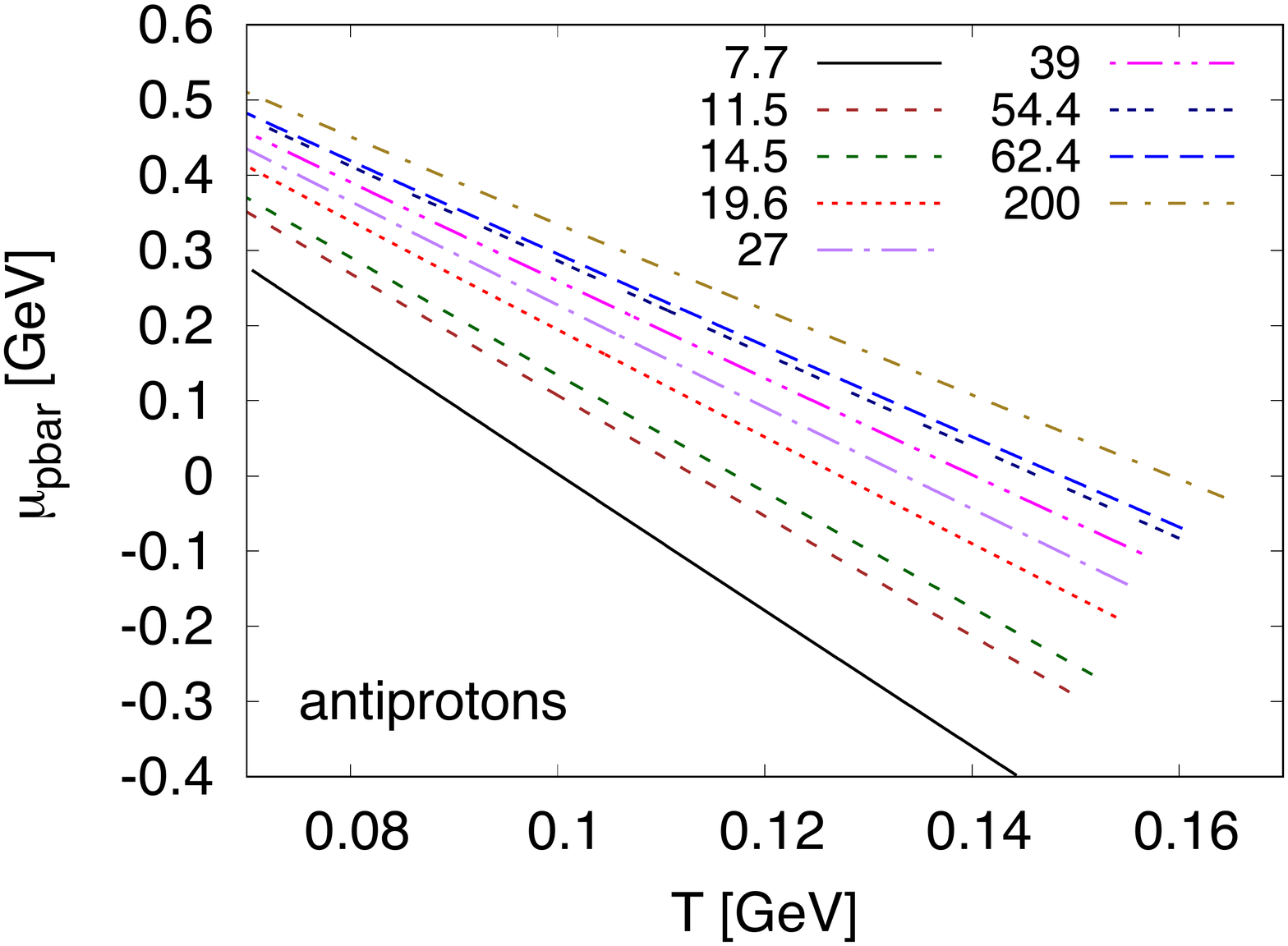}
\end{center}
\caption{Temperature dependence of the  chemical potentials for  protons (upper panel) and antiprotons
(lower panel), for central Au+Au reactions at different collision energies from $\sqrt{s_{NN}}=7.7$ GeV to $\sqrt{s_{NN}}=200$ GeV as indicated in the legend.}
\label{f:mus} 
\end{figure}

The calculated temperature dependences of the chemical potentials for protons and antiprotons are shown in Fig.~\ref{f:mus} for central Au+Au reactions at various collision energies in the RHIC-BES energy regime.
%
%%%%%%%%%%%%%%%%%%%%%%%%%%%
%

%%%%%%%%%%%%%%%%%%%%%%%%%%%
\subsection{Baryon number fluctuations}
\label{ss:bfluct}

Ideally, of  interest are fluctuations of the baryon number. Thus, we need to inspect them in order to see
how they evolve with decreasing temperature and how they can be related to the measurable net-proton number fluctuations. 

Unfortunately, out of equilibrium one can not define a baryochemical potential, thus the cumulants cannot be expressed
shortly via one derivative of the complete partition function. Hence, the cumulants have to be calculated by an extension 
of Eq.~(\ref{e:cumppbar})
\begin{equation}
\cum{(\Delta B)}{l} = \sum_R B_R^l \cum{(\delN_R)}{l}\,  ,
\label{e:Bcums}
\end{equation}
where the sum counts all hadron species and resonances, and $B_R$ is the baryon number of species $R$.

%%%%%%%%%%%%%%%%%%%%%%%%%%%
\subsection{Summary of the model}
\label{ss:msum}

Cumulants of the proton number distribution are calculated from Eqs.~(\ref{e:allpcums}) and their 
volume-independent ratios 
via the relations~(\ref{e:volind}). Analogous results for the baryon number fluctuations are obtained through 
Eq.~(\ref{e:Bcums}). 

The cumulants of the numbers of different resonance species are determined through Eqs.~(\ref{e:meanR})
and (\ref{e:cumR}), with the chemical potentials calculated in Section~\ref{ss:PCE}.

%%%%%%%%%%%%%%%%%%%%%%%%%%%%%%%%%%%%%%%%%%%%%%%%%%%%%%%%%%%
\section{Results}
\label{s:results}
\subsection{Net protons}

%%%%%%%%%%%%%%%%%%%%%%%%%%%
\begin{figure*}[ht!]
\begin{center}
	\includegraphics[width=0.7\textwidth]{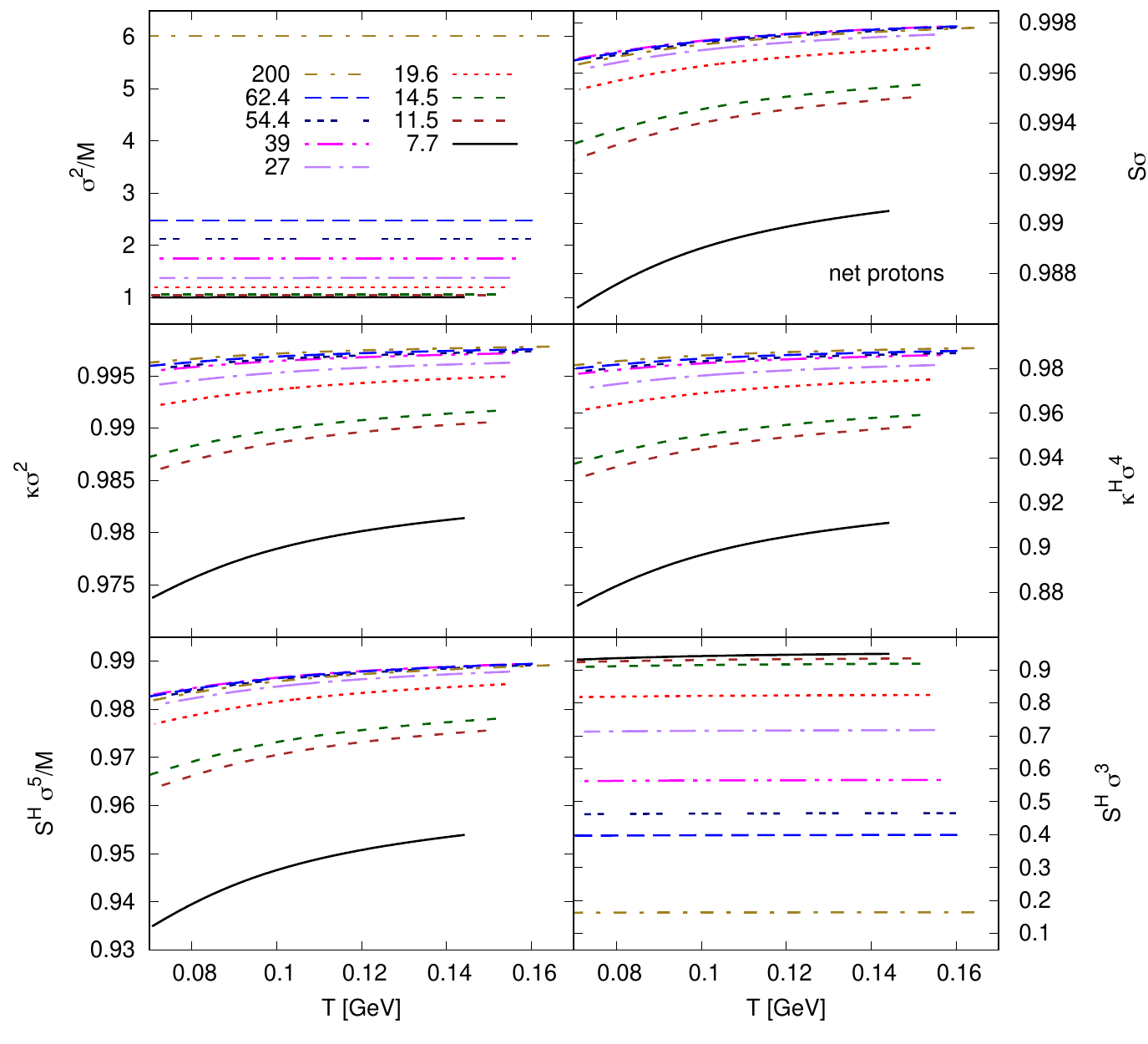}
\end{center}
\caption{Volume-independent ratios of net-proton number cumulants as functions of temperature, for  central Au+Au reactions at different collision energies as indicated in the Figure. }
\label{f:pflu} 
\end{figure*}
%%%%%%%%%%%%%%%%%%%%%%%%%%%
Now we are in the position to answer the question of the evolution of the net-proton number susceptibility (cumulant) ratios from chemical freeze-out to the final decoupling of the system.

In Fig.~\ref{f:pflu} we present the temperature dependence of cumulants for the net-proton number distribution, calculated
for different collision energies of the RHIC-BES. From top to bottom, we show the ratios of the cumulants in increasing order. The largest variation between chemical freeze-out and the temperature of 70~MeV is 
for higher cumulants and lower collision energies: $\cum{(\delN_{p-\bar p})}{6}/\cum{(\delN_{p-\bar p})}{2} = \kappa^H\sigma^4$ decreases by about 
4\%. 
 
%Recall that PCE is formulated by keeping the mean of the proton number distribution constant. Yet we see that there is barely any 
%change for all cumulant ratios. 

One also notes that the ratios of odd-to-second order cumulants tend to be smaller than 1 and 
decrease considerably as $\sqrt{s_{NN}}$ reaches the highest values in our energy scan. Inversely to that, the ratios of even-to-first 
order cumulants are above 1 and increase largely as $\sqrt{s_{NN}}$ approaches 200~GeV. How can we understand
\begin{enumerate}[(a)]
    \item the apparent flatness of the curves, i.e. the very weak temperature dependence and
    \item the peculiar relations of the odd and even cumulant ratios?
\end{enumerate}

Let us start with the second question. This is straightforwardly understood from 
Eq.~(\ref{e:cumppbar}): for odd orders the antiproton term is subtracted from that of protons, while for even orders they are added together.

The apparent flattness shows up because all cumulants are identical in Boltzmann statistics. Then, by fixing the mean in PCE, all higher cumulants are nearly fixed, as well. The differences are due to departure of the appropriate quantum-statistical distribution
from the classical Boltzmann one.

Let us understand it more deeply. 
When discussing Eqs.~(\ref{e:meanR}) and (\ref{e:cumR}), we mentioned that 
the sums may be limited to just a few terms. We now apply the Boltzmann approximation and only keep the first term. Then, cumulants of 
all orders are equal,
\begin{equation}
\cum{(\delN_R)}{l} = \frac{g_R V}{2\pi^2} m_R^2 T \exp\left ( \frac{\mu_R}{T}\right ) K_2\left ( \frac{m_R}{T} \right ) = \cum{N_R}{}\, .
\end{equation}
This universal relation can subsequently be inserted in eqs.~(\ref{e:allpcums}). All the terms which multiply cumulants of different orders 
can then be summed up, and one recognises that they all together give just $\pr$
\begin{equation}
\label{e:csres}
\cum{(\delN_p)}{l} = \sum_R \pr \cum{N_R}{} = \cum{N_p}{}\,  .
\end{equation}
Since the first moment $ \cum{N_p}{}$ is temperature-independent by construction in PCE, all the higher moments are, as well constant in the Boltzmann approximation. Hence, 
the dominant contribution to the ratios should not show a temperature dependence. The weak observed temperature dependence is generated solely by quantum-statistical effects.  Thus, in PCE we see that we basically can access the moments  as they  are set at the chemical freeze-out even by 
measuring protons that come out from a fireball that cools further down. 

%%%%%%%%%%%%%%%%%%%%%%%%%%

\subsection{Net baryons}

%%%%%%%%%%%%%%%%%%%%%%%%%%%
\begin{figure*}[ht!]
\begin{center}
	\includegraphics[width=0.7\textwidth]{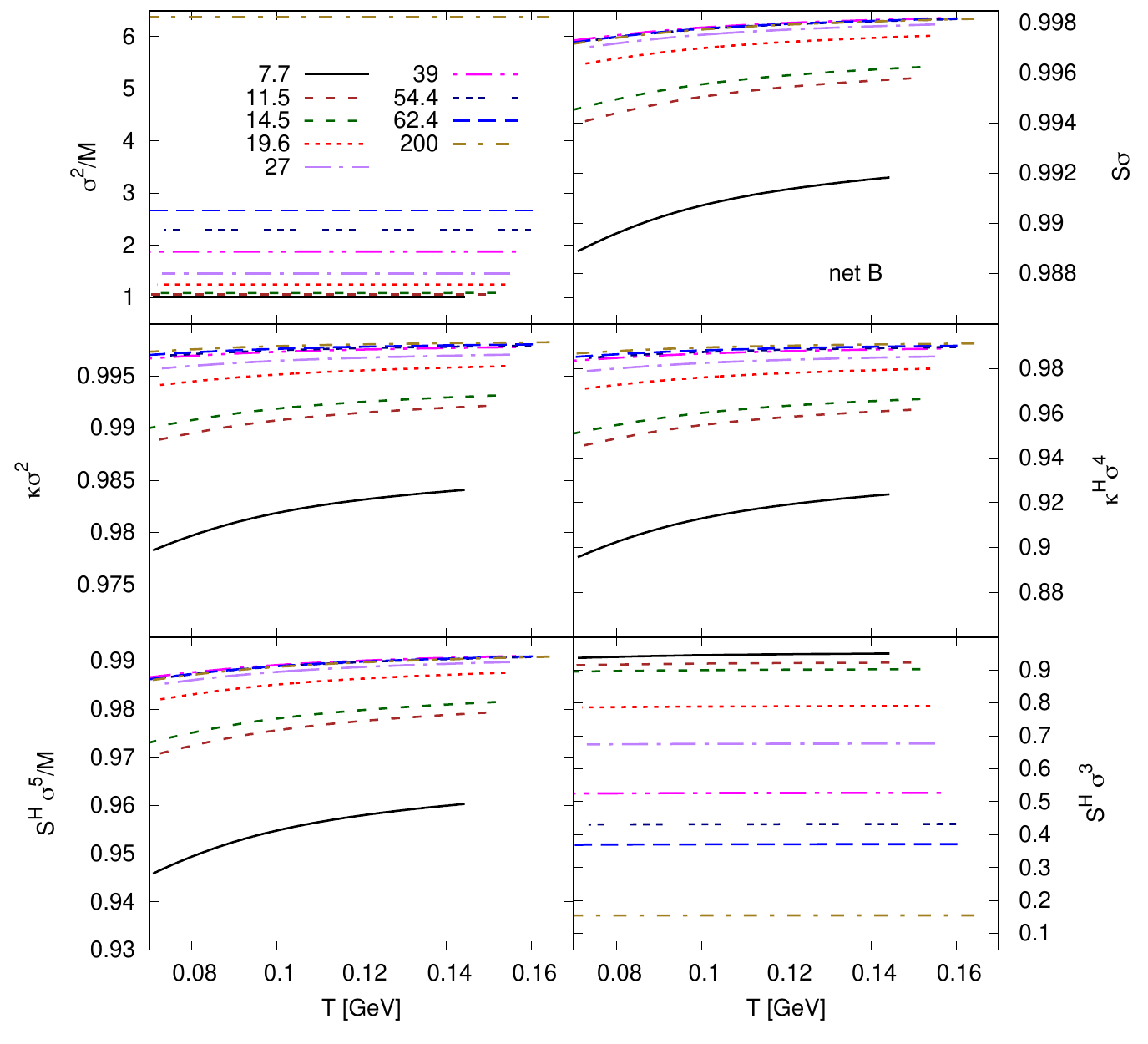}
\end{center}
\caption{Volume-independent ratios of net-baryon number cumulants as functions of temperature, for  central Au+Au reactions at different collision energies as indicated in the Figure. }
\label{f:bnum} 
\end{figure*}
%%%%%%%%%%%%%%%%%%%%%%%%%%%

Next, we look at the fluctuations of the baryon number. The cumulant ratios as functions of temperature are depicted in Fig.\ref{f:bnum}.
The ratios are very similar to those of net-proton number. In fact, this is not entirely trivial to expect because the moments of the proton number 
distribution are influenced by the randomness of the resonance decays, while there is no such contribution when the baryon number is 
studied. Once again, the key to understanding the feature is in realising that in the Boltzmann approximation, for cumulants at any order 
contributions due to resonance decays add up to the universal term $\pr\cum{N_R}{}$, as discussed in relation to Eq.~(\ref{e:csres}).

%%%%%%%%%%%%%%%%%%%%%%%%%%%
\begin{figure}[ht!]
\begin{center}
	\includegraphics[width=0.45\textwidth]{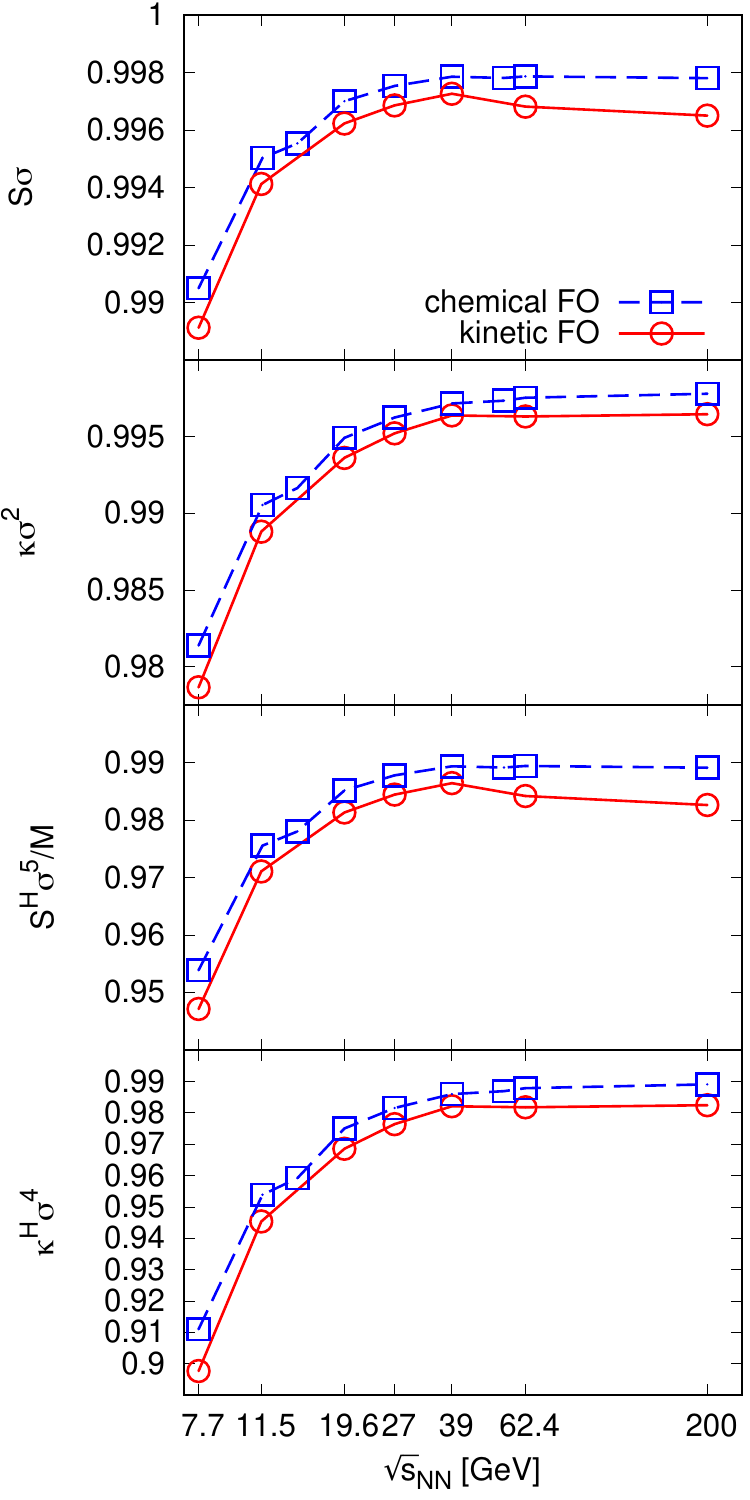}
\end{center}
\caption{Values of the volume-independent ratios of net-proton number for central Au+Au reactions at chemical freeze-out (squares) and at the kinetic freeze-out
(circles) according to \cite{Melo:2019mpn}.}
\label{f:fo} 
\end{figure}
%%%%%%%%%%%%%%%%%%%%%%%%

\subsection{Energy dependence}
We further study the effects of cooling within the PCE model on the values of cumulant ratios, as they are observed experimentally. 
It is clear that there is a tension between the statistical model and the observed data which indicate a strong enhancement of $\kappa\sigma^2$
as the collision energy reaches down to the 7.7~GeV per colliding $NN$ pair. The collision energy dependence of selected ratios is plotted in
Fig.~\ref{f:fo}. With lowering the energy, there is always only a small decrease of all the studied ratios 
evaluated at chemical freeze-out. It becomes more pronounced 
with the higher order moments, getting to almost 10\% for $\kappa^H\sigma^4$. The values are then slightly lowered if also 
cooling and PCE down to kinetic freeze-out 
is taken into account. Again, the decrease is largest for $\kappa^H\sigma^4$, where it is about 1\%. 

\subsection{Resonance spectrum}
%%%%%%%%%%%%%%%%%%%%%%%%%%%
\begin{figure}[t]
\begin{center}
	\includegraphics[width=0.45\textwidth]{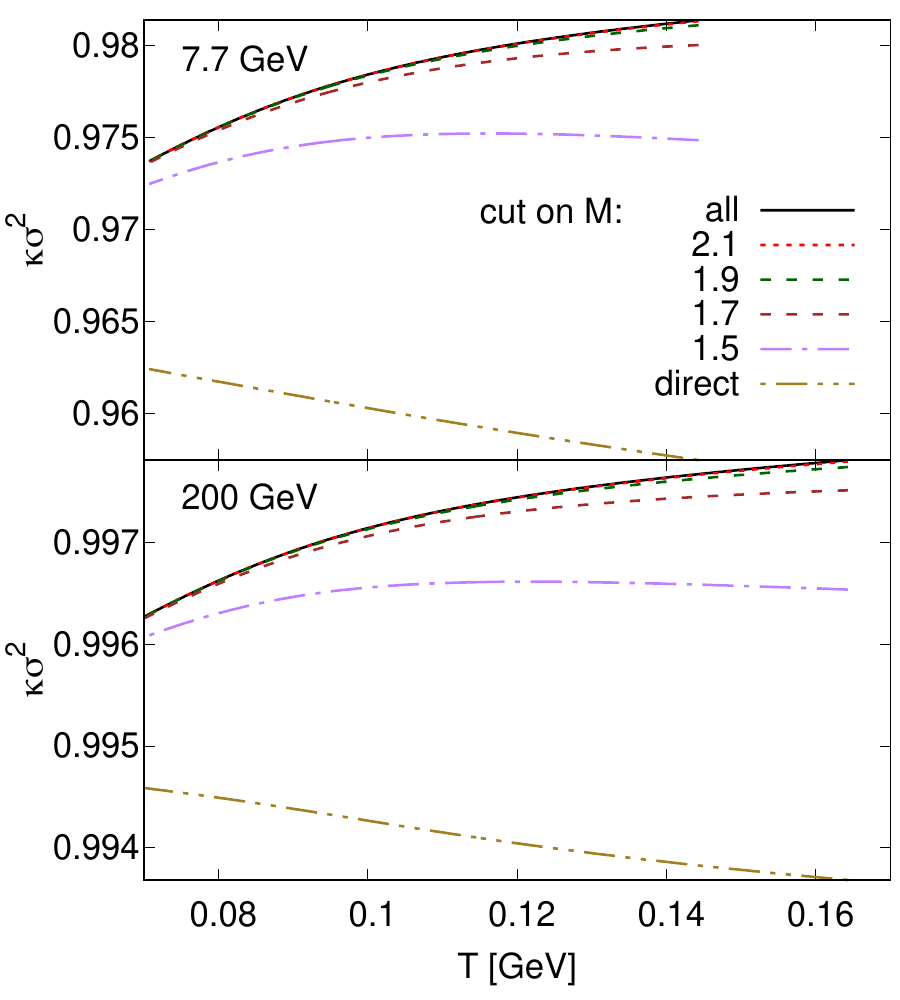}
\end{center}
\caption{The ratio $\kappa\sigma^2$ as function of temperature for central Au+Au reactions at $\sqrt{s_{NN}} = 7.7$~GeV (upper panel) and 200 GeV (lower panel). Different curves show results where the number of resonances included into calculation was constrained by an upper cut on the resonance mass. The curves range from no resonances included (dash-double-dotted curves) up to all resonances included (solid curves).}
\label{f:Mcut} 
\end{figure}
%%%%%%%%%%%%%%%%%%%%%%%%
Cumulants of proton and antiproton number distributions are calculated in Eqs.~(\ref{e:allpcums}) by summing contributions 
to (anti)proton production from decays of many sorts of resonances. One may then wonder how important the individual 
contributions are to the overall result. To explore this question, as an example, in Fig.~\ref{f:Mcut} we investigate the dependence 
of $\kappa\sigma^2$ on the temperature, where we limit the mass spectrum of resonances included into the calculation. We pick this 
ratio of cumulants, because the higher-order cumulants are more sensitive to the precision of the calculation and so may be more 
influenced by the cutoff in the inclusion of higher-mass resonances. Nevertheless, for the illustration we have not chosen the 
fifth and sixth-order cumulant ratios, since they are not yet measured at all energies. Calculations for two collision energies are done from the extremes 
of the interval that we investigate: 7.7 and 200 GeV per $NN$ pair. One can observe that limiting calculations to only direct proton production 
even leads to qualitatively incorrect conclusions: with lowering the temperature the ratio grows instead of decreasing. Its value is smaller than 
that of the full calculation and the largest deviation of about 2\% is 
there at low $\sqrt{s_{NN}}$. Just the inclusion of $\Delta$ and $N(1440)$
resonances improves the ratio considerably. Including all resonances up to the mass of 2.1~GeV practically yields the same results as the full 
calculation. 

%%%%%%%%%%%%%%%%%%%%%%%%%%%

\subsection{Kaons}

%%%%%%%%%%%%%%%%%%%%%%%%%%%
\begin{figure*}
\begin{center}
	\includegraphics[width=0.7\textwidth]{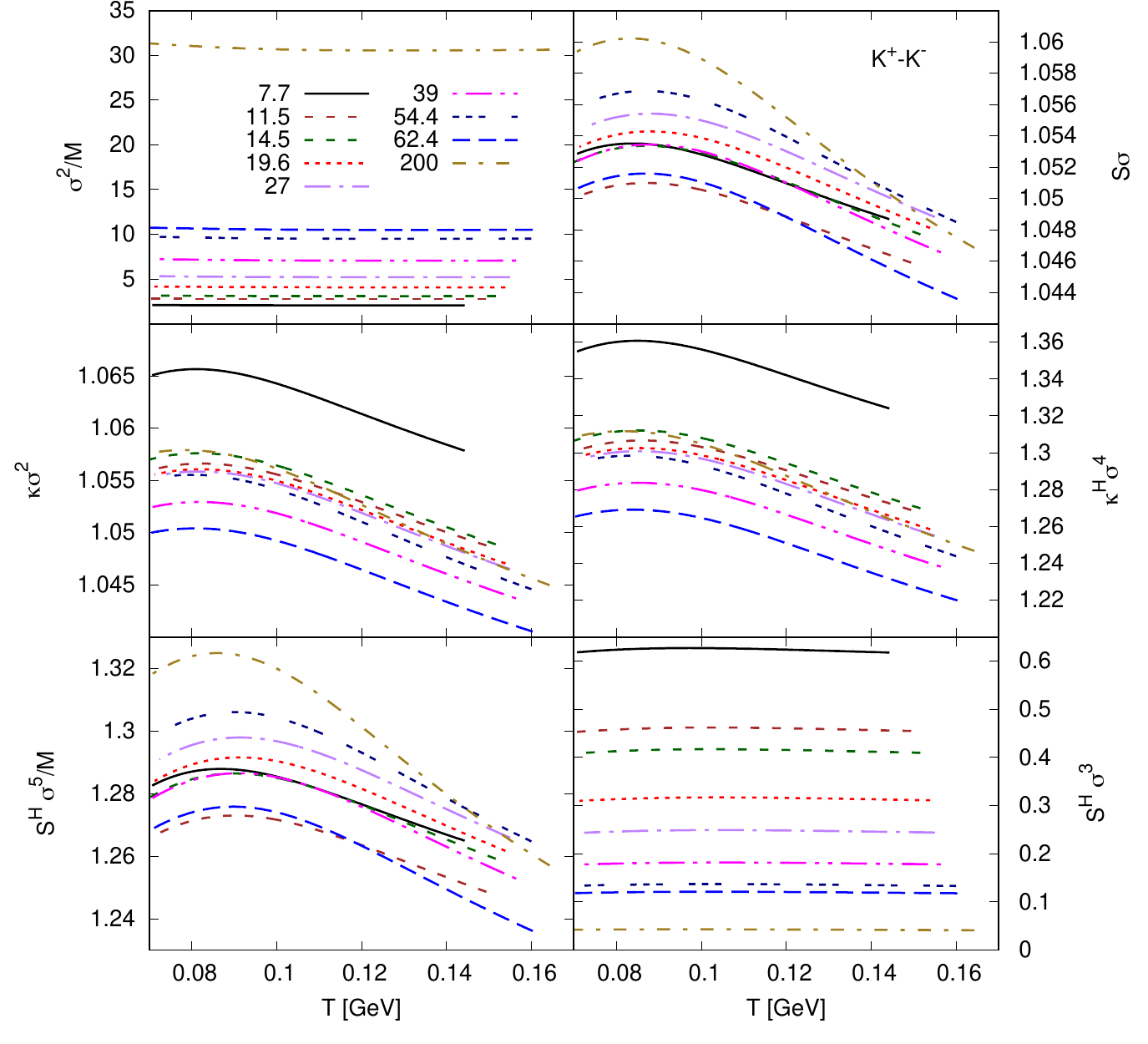}
\end{center}
\caption{Volume-independent ratios of the $(K^+ - K^-)$ number cumulants as functions of temperature, for central Au+Au reactions at different collision energies as indicated in the Figure. }
\label{f:kaons} 
\end{figure*}
%%%%%%%%%%%%%%%%%%%%%%%%%%%
Finally, we have discussed above that the temperature dependence of the cumulant ratios is non-constant only as a result of the difference between the
Boltzmann and the appropriate quantum-statistical distribution. Due to the high proton mass as compared to the temperature scale, 
that difference is rather small. However, it might be larger for particles with lower mass. Hence, we look at the cumulants of the 
($K^+-K^-$) number distribution. 
The same formalism may be used as we did for protons, because there is no resonance that would 
produce two $K^+$'s or $K^-$'s. 
The only caveat is that we had to leave out the decays of $\phi$ meson 
from kaon production. Since it may decay into a $K^+K^-$ pair, it 
introduces correlation between $K^+$ and $K^-$ yields. However, it 
does not change the difference of $K^+-K^-$ multiplicities, and 
so is justifiably left out. 
Having the list of resonances, the calculation is  straightforward and we show the results in 
Fig.~\ref{f:kaons}. Numerically, the results are different from those for the net-proton number distribution, but the change of the ratios 
due to the decrease of the temperature is not bigger than 5\% 
even in the most `extreme' case of $S^H\sigma^5/M$, In contrast to the net-proton 
number fluctuations, the cumulant ratios of the $K^+-K^-$ number distribution generally increases when the temperature is lowered
from its chemical freeze-out value to the thermal freeze-out.

%%%%%%%%%%%%%%%%%%%%%%%%%%%%%%%%%%%%%%%%%%%%%%%%%%%%%%%%%%%
\section{Conclusions}
\label{s:conc}

We have analyzed the evolution of ratios of net-proton number cumulants and net-kaon number cumulants from the chemical freeze-out to the kinetic freeze-out. To this aim, we employed a hadron resonance gas model with the assumption of partial chemical equilibrium.

We found only a weak dependence of the cumulant ratios when decreasing the temperature while keeping the partial chemical equilibrium, 
as it is dictated by the observed average-hadron abundance ratios. Nevertheless these improved statistical model estimates provide a better baseline for the comparison of theory to data and we have demonstrated that 
the conclusions based on cumulants calculated at chemical freeze-out remain valid after cooling. 

As discussed in the introduction, further effects may also contribute to the baseline prediction which are not included in our model:
\begin{enumerate}
\item The grand-canonical 
formalism behind the hadron resonance gas model does not include the total baryon number conservation, which 
has been worked out in detail recently
\cite{Braun-Munzinger:2020jbk,Vovchenko:2020tsr,Vovchenko:2020gne}.
\item Also, the impact of limited acceptance and effectivity of the detector, for 
which a formalism has been published recently \cite{Braun-Munzinger:2020jbk}, was not included here.
These first two effects are interconnected, 
because the limited acceptance 
introduces fluctuations by looking at a subset of a larger sample of baryons with fixed total baryon number. Nevertheless, we can expect that 
the impact due to those two effects on our results would be similar as at the chemical freeze-out. Tiny modifications may be expected 
due to narrower momentum distribution of hadrons at lower temperature and a smaller proportion of protons coming from resonance 
decays. 

\item Full chemical equilibrium is assumed between one sort of stable hadrons and all heavier resonance states
which may decay into that hadron. For example, protons, $\Delta$'s, $N(1440)$'s etc are always in relative chemical equilibrium. 
In contrast to them, in order to keep the total produced number of stable hadrons constant, no processes are allowed which would 
modify their numbers, e.g.~$K^+\Lambda \leftrightarrow p\pi^0$ is assumed not to run. The particularly striking consequence is then that 
also $p\bar p$ annihilation does not exist in this model.
This is a strong and simplifying assumption made in the present the model, which must be remembered
when interpreting these results.

\item Finally, to keep track of the expanding volume, the total entropy is assumed to remain constant during the expansion. This is also rather 
simplifying assumption which might be revised. 

\item Note also that recently the whole concept of the hadron resonance gas was put under scrutiny. The representation of interactions
through the presence of resonances is imprecise and has been replaced by a more direct treatment with the help of S-matrix. 
So far, this approach has been used in the description of mean numbers of particles \cite{Cleymans:2020fsc}. The question
remains open, which consequences this improved formulation of the model has on the higher order cumulants. 
\end{enumerate}

Nevertheless, in spite of all potential shortcomings of the model used in our study, we expect that our reasoning would stay 
unchanged, that there is only a modification on the level of per cent of the volume-independent cumulant ratios in 
cooling fireball after the chemical freeze-out. 

%%%%%%%%%%%%%%%%%%%%%%%%%%%%%%%%%%%%%%%%%%%%%%%%

\begin{acknowledgments}
The collaboration has been supported by the COST Action CA15213 (THOR). 
BT acknowledges support by the project Centre of Advanced Applied Sciences with number CZ.02.1.01/0.0/0.0/16-019/0000778, which is co-financed by the European Union. BT acknowledges support from VEGA 1/0348/18. PH acknowledges support by the Frankfurt Institute of Advanced Studies (FIAS).
\end{acknowledgments}

\appendix

%%%%%%%%%%%%%%%%%%%%%%%%%%%%%%%%%%%%%%%%%%%%%%%%%%%%%%%%%%%
\section{Determination of the proton number cumulants}
\label{s:app}

%%%%%%%%%%%%%%%%%%%%%

\subsection{Fixed number of resonances}

We shall start our derivation from the proton production by a single species of baryon resonances with a  fixed number $N_R$. (There are no protons in decays of mesonic resonances.) 

Due to baryon number conservation, at most one proton can 
be produced in a decay of a resonance with mass smaller than three times the proton mass. A resonance can decay via various channels. Some of them are accounted for here as chain decays, so a channel is understood to sum up all stable hadrons. 
Since there is either 0 or 1 proton in the final state of the decay, the probability that there is a proton after the decay equals the 
mean number of protons from the decay
\begin{equation}
\pr = \sum_{r} b_r n_{r}\,  ,
\end{equation}
where the sum goes through all decay channels (including all subsequent chain decays), $b_r$ is the probability (branching ratio) of 
a certain decay channel $r$ and $n_r$ is the number of protons produced in the specific channel $r$ (either 0 or 1). 

Therefore, the probability to produce $N$ protons by the decays of $N_R$ resonances is given by a binomial distribution
\begin{equation}
P(N;N_R) = 
\left (\!\! \!\begin{array}{c}N_R\\N\end{array}\!\!\!\right ) \pr^N (1-\pr)^{N_R-N}\,  .
\label{e:bind}
\end{equation}

Using Eq.~(\ref{e:cumder}) one derives the corresponding cumulant-generating function as
\begin{eqnarray}
K_b(i\xi) &= & \ln \left \{ \sum_{N=0}^{N_R}  e^{i\xi N} \left (\!\! \!\begin{array}{c}N_R\\N\end{array}\!\!\!\right ) \pr^N (1-\pr)^{N_R-N}
\right \}
\nonumber \\
& = & \ln \left \{ \sum_{N=0}^{N_R}   \left (\!\! \!\begin{array}{c}N_R\\N\end{array}\!\!\!\right ) e^{i\xi N}\pr^N (1-\pr)^{N_R-N}
\right \} 
\nonumber \\ &=& 
\ln \left \{ \sum_{N=0}^{N_R}   \left (\!\! \begin{array}{c}N_R\\N\end{array}\!\!\right ) (e^{i\xi}\pr)^N (1-\pr)^{N_R-N}
\right \}
\nonumber \\
& = & \ln \left ( e^{i\xi}\pr + (1-\pr)\right )^{N_R}
\nonumber \\
& = & N_R \ln \left ( e^{i\xi}\pr + (1-\pr)\right )\,  .
\label{e:cgfb}
\end{eqnarray}

%%%%%%%%%%%%%%%%%%%%%%%%%%

\subsection{Single sort of resonances with fluctuating number}

In a grand-canonical system the number of resonances fluctuates. Let us, for the moment, assume that the probability to 
have $N_R$ resonances is $P_R(N_R)$, which is properly normalised
\[
\sum_{N_R = 0}^{\infty} P_R(N_R) = 1\,  .
\]

To get the probability that $N$ protons appears, we have to sum up over all $N_R$ which fulfil $N_R \ge N$
\begin{equation}
P(N) = \sum_{N_R=N}^{\infty} P_R(N_R) P(N;N_R)\,  ,
\end{equation}
where $P(N;N_R)$ is the binomial distribution, as in Eq.~(\ref{e:bind}). 

The cumulant-distribution function, according to definition (\ref{e:cumder}), is 
\begin{eqnarray}
K(i\xi) & = & \ln \left \{ \sum_{N=0}^{\infty} e^{i\xi N} P(N) \right \}
\nonumber \\  
& = & \ln \left \{ \sum_{N=0}^{\infty} e^{i\xi N} \sum_{N_R=N}^{\infty} P_R(N_R) P(N;N_R)\right \} 
\nonumber \\
& = & \ln \left \{ \sum_{N=0}^{\infty}  \sum_{N_R=N}^{\infty}e^{i\xi N} P_R(N_R) P(N;N_R)\right \} 
\,  .
\end{eqnarray}
The trick is now to switch the order of summations
\begin{multline}
K(i\xi) = \ln \left \{ \sum_{N_R=0}^{\infty} \sum_{N=0}^{N_R}  e^{i\xi N} P_R(N_R) P(N;N_R)\right \}  
\\
= 
\ln \left \{ \sum_{N_R=0}^{\infty} P_R(N_R) \sum_{N=0}^{N_R}  e^{i\xi N}  P(N;N_R) \right \} \,  .
\end{multline}
We know how to do the second summation, because this is exactly what was done in deriving Eq.~(\ref{e:cgfb}). We arrive at 
\begin{equation}
\label{e:cgf1}
K(i\xi) = \ln  \left \{ \sum_{N_R=0}^{\infty} P_R(N_R)    \left ( e^{i\xi}\pr + (1-\pr)\right )^{N_R} \right \} \,  .
\end{equation}

From this cumulant-generating function we can now derive the cumulants of the proton number distribution by taking derivatives. 
To simplify the notation and the calculation, let us introduce some shorthands. The argument of the logarithm shall be denoted
\begin{equation}
M = M(\xi) = \sum_{N_R=0}^{\infty} P_R(N_R)    \left ( e^{i\xi}\pr + (1-\pr)\right )^{N_R}\,  ,
\end{equation}
where it is easy to see that 
\begin{equation}
M(0) = 1\, .
\end{equation}
We will also need derivatives of $M$. (We assume that the function is as many times differentiable, as we need.)
\begin{eqnarray}
M' & \equiv &
\frac{\dif M(\xi)}{\dif (i\xi)} 
\nonumber \\
& = & \sum_{N_R=0}^{\infty} P_R(N_R)  N_R \pr 
\nonumber \\
&& \quad \qquad \times \left ( e^{i\xi}\pr + (1-\pr)\right )^{N_R-1} \eix\\
M'(0) & = & \pr  \sum_{N_R=0}^{\infty} P_R(N_R)  N_R = \pr \cum{N_R}{} \,  .
\end{eqnarray}
Higher derivatives will follow. Since factorial moments of the distribution $P_R(N_R)$  will appear frequently, we introduce a notation
\begin{equation}
\label{e:fmomnot}
F_i = \nrfm{i}\,  ,
\end{equation}
where, $F_1 = \langle N_R \rangle$.

Using these shorthands, the derivatives of $M$ evaluated at $\xi=0$ are
\begin{subequations}
\label{e:eMs}
\begin{eqnarray}
M''(0) & = & \pr^2 F_2 + \pr F_1\, ,
\\
M^{(3)}(0) & = & \pr^3 F_3 + 3 \pr^2 F_2 + \pr F_1\, ,
\\
M^{(4)}(0) & = & \pr^4 F_4 + 6\pr^3 F_3 + 7\pr^2 F_2 + \pr F_1 \,  ,
\\
M^{(5)}(0)  & =& \pr^5 F_5 + 10\pr^4 F_4 + 25\pr^3 F_3 + 15 \pr^2 F_2 
\nonumber \\ &&
{} + \pr F_1\,  ,
\\
M^{(6)}(0) & = &  \pr^6 F_6 + 15\pr^5 F_5 + 65 \pr^4 F_4 + 90 \pr^3 F_3 
\nonumber \\ && 
{} + 31 \pr^2 F_2 + \pr F_1\,  .
\end{eqnarray}
\end{subequations}
These shorthands are then used in the derivatives of the cumulant-generating function. 
We also need the relations between factorial moments and cumulants, which are summarised in Eqs.~(\ref{e:fc}). 
The cumulants of the proton number distribution are then calculated via Eq.~(\ref{e:cumdef}) with the cumulant-generating 
function defined in Eq.~(\ref{e:cgf1}). We obtain
\begin{widetext}
\begin{subequations}
\label{e:cumsviaM}
\begin{eqnarray}
\cum{N}{} & = & \frac{M'(0)}{M(0)} = M'(0) \,  ,
\\
\cum{(\Delta N)}{2} & = & M''(0)M(0) - (M'(0))^2\, ,
\\
\cum{(\Delta N)}{3} & = & M^{(3)}(0)  - 3 M''(0) M'(0)  + 2(M'(0))^3\,  ,
\\
\cum{(\Delta N)}{4} & = & M^{(4)}(0)  - 4M^{(3)}(0) M'(0)
 + 12 M''(0) (M'(0))^2 - 3 (M''(0))^2 - 6(M'(0))^4 \,  ,
\\
\cum{(\Delta N)}{5} & = &\Md{5}(0)   -  5 \Md{4}(0) M'(0)   -  10 \Md{3}(0) M''(0)   
+ 20 \Md{3}(0) (M'(0))^2 - 60 M''(0) (M'(0))^3 
\nonumber \\ &&  {} + 30 (M''(0))^2 M'(0)  + 24 (M'(0))^5 \, ,
\\
\cum{(\Delta N)}{6} & = & \Md{6}(0)-   6 \Md{5}(0)M'(0) - 15 \Md{4}(0)M''(0) + 30 \Md{4}(0) (M'(0))^2  - 10 (\Md{3}(0))^2
 \nonumber \\ &&  {}  + 120 \Md{3}(0) M''(0) M'(0) - 120 \Md{3}(0) (M'(0))^3  - 270 (M''(0))^2 (M'(0))^2
\nonumber \\ &&  {}  + 30 (M''(0))^3   + 360 M''(0)(M'(0))^4 - 120 (M'(0))^6
\,  .
\end{eqnarray}
\end{subequations}

The last steps are tedious but straightforward. First, the derivatives of $M$ are expressed via Eqs.~(\ref{e:eMs}) and inserted into 
Eqs.~(\ref{e:cumsviaM}). Then, the factorial moments in those expressions are rewritten in terms of the cumulants of the resonance 
 number distribution, which are derived in Eqs.~(\ref{e:fc}). This leads to the following expressions for the proton number cumulants
 \begin{subequations}
\label{e:1pcums}
\begin{eqnarray}
\cum{N_p}{} & = &  \pr \cum{N_R}{}\, ,  \\
\cum{(\delN_p)}{2} & = & \ \pr^2 \cum{(\delN_R)}{2} + \pr(1-\pr) \cum{N_R}{} \, , \\
\cum{(\delN_p)}{3} & = &   \pr^3 \cum{(\delN_R)}{3}  + 3 \pr^2(1-\pr)  \cum{(\delN_R)}{2} 
+ \pr(1-\pr)(1-2\pr)\cum{N_R}{}  \,  , \\
\cum{(\delN_p)}{4} & = & 
\pr^4\cum{(\delN_R)}{4} + 6\pr^3(1-\pr)\cum{(\delN_R)}{3}
 + \pr^2(1-\pr)(7-11\pr)\cum{(\delN_R)}{2} 
 \nonumber \\
 &&  {}
+ \pr(1-\pr)(1-6\pr+6\pr^2) \cum{N_R}{}  \,  , \\
\cum{(\delN_p)}{5} & = & 
\pr^5\cum{(\delN_R)}{5} +10 \pr^4(1 - \pr)  \cum{(\delN_R)}{4} 
+ 5 \pr^3 (1-\pr)( 5-7\pr) \cum{(\delN_R)}{3}
\nonumber \\ &&  {}
+5 \pr^2 (1-\pr)(10\pr^2 - 12\pr+3) \cum{(\delN_R)}{2} 
\nonumber \\ &&  {}
+  \pr(1 - \pr)  (1 - 2 \pr)  (12\pr^2 - 12 \pr  +1)  \cum{N_R}{} \, ,
\\
\cum{(\delN_p)}{6} & = & 
\pr^6\cum{(\delN_R)}{6} +15\pr^5(1-\pr)\cum{(\delN_R)}{5}+5\pr^4(1-\pr)(13-17\pr)\cum{(\delN_R)}{4}
\nonumber \\ 
&&  {}
+ 15\pr^3(1-\pr)(15\pr^2-20\pr+6)\cum{(\delN_R)}{3} 
\nonumber \\ 
&&  {}
- \pr^2(1-\pr)(  274 \pr^3 -476 \pr^2  +239 \pr - 31  )\cum{(\delN_R)}{2}
\nonumber \\ 
&& {}
+ \pr(1-\pr)( 120 \pr^4 - 240 \pr^3 + 150 \pr^2  - 30 \pr + 1)\cum{N_R}{} \,  .
\end{eqnarray}
\end{subequations}
\end{widetext}
Since cumulants are additive for random numbers that are added together, if protons are produced from decays of many 
sorts of resonances, we have to sum the above expressions through all of them. This leads to  Eqs.~(\ref{e:allpcums}).

%%%%%%%%%%%%%%%%%%%%%%%%%%%%%%%%%%%%%%%%%%%%%%%%%%%%%%%%%%%

\section{Relations between factorial moments, cumulants, and moments}
\label{s:app2}

Here, we review the relations which are useful if one set of statistics is needed to be replaced by another one. These relations 
are generic and can be derived by taking derivatives of the \emph{characteristic function}
\[
\varphi(i\xi) = \sum_{N=0}^{\infty} e^{i\xi N} P(N)\,  ,
\] 
and the cumulant-generating function defined in Eq.~(\ref{e:cumder})
\[
K(i\xi) = \ln [\varphi(i\xi)]\,  .
\]
The moments are obtained as 
\begin{equation}
\mom{m} \equiv \langle N^m\rangle = \left . \frac{\dif^m \chf}{\dif(i\xi)^m} \right |_{\xi=0} \,  ,
\label{e:md}
\end{equation}
and the cumulants are obtained as 
\begin{equation}
\cum{(\Delta N)}{m} =  \left . \frac{\dif^m K(i\xi)}{\dif(i\xi)^m} \right |_{\xi=0}\,  .
\label{e:b2cd}
\end{equation}

Systematically applying the derivatives in Eq.~(\ref{e:b2cd}) and using Eq.~(\ref{e:md}) to express them with the help of the moments
$\mom{m}$ yields
\begin{widetext}
\begin{subequations}
\begin{eqnarray}
\cum{N}{} & = & \mom{1} \,  ,
\\
\cum{(\Delta N)}{2} & = & \mom{1}^2+\mom{2} \,  ,
\\
\cum{(\Delta N)}{3} & = & 2 \mom{1}^3-3 \mom{1} \mom{2}+\mom{3} \,  ,
\\
\cum{(\Delta N)}{4} & = & -6 \mom{1}^4+12 \mom{1}^2 \mom{2}
-3 \mom{2}^2-4 \mom{1} \mom{3}+\mom{4} \,  ,
\\
\cum{(\Delta N)}{5} & = & 24 \mom{1}^5-60 \mom{1}^3 \mom{2}+30 \mom{1} \mom{2}^2
+20 \mom{1}^2 \mom{3}
-10 \text{$\mu$2} \mom{3}-5 \mom{1} \mom{4}
+\mom{5} \,  ,
\\
\cum{(\Delta N)}{6} & = & 
-120 \mom{1}^6+360 \mom{1}^4 \mom{2}-270 \mom{1}^2 \mom{2}^2
+30 \mom{2}^3-120 \mom{1}^3 
\mom{3}+120 \mom{1} \mom{2} \mom{3}
-10 \mom{3}^2+30 \mom{1}^2 \mom{4}
\nonumber \\ && {}
-15 \mom{2} \mom{4}-6 \mom{1} \mom{5}+\mom{6} \,  .
\end{eqnarray}
\end{subequations}
These relations can be inverted
\begin{subequations}
\label{e:mk}
\begin{eqnarray}
\mom{1} & = & \cum{N}{} \,  ,
\\
\mom{2} & = & \cum{N}{}^2+\cum{(\Delta N)}{2} \,  ,
\\
\mom{3} & = & \cum{N}{}^3+3 \cum{N}{} \cum{(\Delta N)}{2}+\cum{(\Delta N)}{3} \,  ,
\\
\mom{4} & = & \cum{N}{}^4+6 \cum{N}{}^2 \cum{(\Delta N)}{2}+3 \cum{(\Delta N)}{2}^2
+4 \cum{N}{} \cum{(\Delta N)}{3}+\cum{(\Delta N)}{4} \,  ,
\\
\mom{5} & = & \cum{N}{}^5+10 \cum{N}{}^3 \cum{(\Delta N)}{2}
+15 \cum{N}{} \cum{(\Delta N)}{2}^2+10 \cum{N}{}^2 \cum{(\Delta N)}{3}
+10 \cum{(\Delta N)}{2} \cum{(\Delta N)}{3}
\nonumber \\ &&  {}
+5 \cum{N}{} \cum{(\Delta N)}{4}
+\cum{(\Delta N)}{5} \,  ,
\\
\mom{6} & = & \cum{N}{}^6+15 \cum{N}{}^4
\cum{(\Delta N)}{2}
+45 \cum{N}{}^2 \cum{(\Delta N)}{2}^2+15 \cum{(\Delta N)}{2}^3
+20 \cum{N}{}^3 \cum{(\Delta N)}{3}
\nonumber \\ &&  {}
+60 \cum{N}{} \cum{(\Delta N)}{2} \cum{(\Delta N)}{3}
+10 \cum{(\Delta N)}{3}^2
+15 \cum{N}{}^2 \cum{(\Delta N)}{4}
+15 \cum{(\Delta N)}{2} \cum{(\Delta N)}{4}
\nonumber \\ &&  {}
+6 \cum{N}{} \cum{(\Delta N)}{5}
+\cum{(\Delta N)}{6} \,  .
\end{eqnarray}
\end{subequations}

\emph{Factorial moments} are defined in Eq.~(\ref{e:fmomnot}),
and can be expressed through the moments as
\begin{subequations}
\begin{eqnarray}
\fm{1} & = & \mom{1} \,  ,
\\
\fm{2} & = & -\mom{1}+\mom{2} \,  ,
\\
\fm{3} & = & 2 \mom{1}-3 \mom{2}+\mom{3} \,  ,
\\
\fm{4} & = & -6 \mom{1}+11 \mom{2}-6 \mom{3}+\mom{4} \,  ,
\\
\fm{5} & = & 24 \mom{1}-50 \mom{2}+35 \mom{3}-10 \mom{4}+\mom{5} \,  ,
\\
\fm{6} & = & -120 \mom{1}+274 \mom{2}-225 \mom{3}+85 \mom{4} \,  .
-15 \mom{5}+\mom{6}
\end{eqnarray}
\end{subequations}

Finally, inserting from Eqs.~(\ref{e:mk}) into these equations we can express factorial moments with the help of the cumulants:
\begin{subequations}
\label{e:fc}
\begin{eqnarray}
\fm{1} & = &\cum{N}{} \,  ,
\\
\fm{2} & = & -\cum{N}{}+\cum{N}{}^2+\cum{(\Delta N)}{2} \,  ,
\\
\fm{3} & = & 2\cum{N}{}+\cum{N}{}^3+3\cum{N}{}\cum{(\Delta N)}{2}-3 \left(\cum{N}{}^2+\cum{(\Delta N)}{2}\right)+\cum{(\Delta N)}{3} \,  ,
\\
\fm{4} & = & -6\cum{N}{}+\cum{N}{}^4+6\cum{N}{}^2\cum{(\Delta N)}{2}+3\cum{(\Delta N)}{2}^2+11 \left(\cum{N}{}^2+\cum{(\Delta N)}{2}\right)+4\cum{N}{}\cum{(\Delta N)}{3}
\nonumber \\ &&  {}
-6 \left(\cum{N}{}^3+3\cum{N}{}\cum{(\Delta N)}{2}+\cum{(\Delta N)}{3}\right)+\cum{(\Delta N)}{4} \,  ,
\\
\fm{5} & = & 24\cum{N}{}+\cum{N}{}^5+10\cum{N}{}^3\cum{(\Delta N)}{2}+15\cum{N}{}\cum{(\Delta N)}{2}^2-50 \left(\cum{N}{}^2+\cum{(\Delta N)}{2}\right)
\nonumber \\ &&  {}
+10\cum{N}{}^2\cum{(\Delta N)}{3}+10\cum{(\Delta N)}{2}\cum{(\Delta N)}{3}+35 \left(\cum{N}{}^3+3\cum{N}{}\cum{(\Delta N)}{2}+\cum{(\Delta N)}{3}\right)
\nonumber \\ &&  {}
+5\cum{N}{}\cum{(\Delta N)}{4}
-10 \left(\cum{N}{}^4+6\cum{N}{}^2
\cum{(\Delta N)}{2}+3\cum{(\Delta N)}{2}^2+4\cum{N}{}\cum{(\Delta N)}{3}+\cum{(\Delta N)}{4}\right)
\nonumber \\ &&  {}
+\cum{(\Delta N)}{5} \,  ,
\\
\fm{6} & = & 
-120\cum{N}{}+\cum{N}{}^6+15\cum{N}{}^4\cum{(\Delta N)}{2}+45\cum{N}{}^2\cum{(\Delta N)}{2}^2+15\cum{(\Delta N)}{2}^3+274 \left(\cum{N}{}^2+\cum{(\Delta N)}{2}\right)
\nonumber \\ &&  {}
+20\cum{N}{}^3\cum{(\Delta N)}{3}+60\cum{N}{}\cum{(\Delta N)}{2}\cum{(\Delta N)}{3}+10\cum{(\Delta N)}{3}^2
\nonumber \\ &&  {}
-225 \left(\cum{N}{}^3+3\cum{N}{}\cum{(\Delta N)}{2}+\cum{(\Delta N)}{3}\right)
+15\cum{N}{}^2
\cum{(\Delta N)}{4}+15\cum{(\Delta N)}{2}\cum{(\Delta N)}{4}
\nonumber \\ &&  {}
+85 \left(\cum{N}{}^4+6\cum{N}{}^2\cum{(\Delta N)}{2}+3\cum{(\Delta N)}{2}^2+4\cum{N}{}\cum{(\Delta N)}{3}+\cum{(\Delta N)}{4}\right)
+6\cum{N}{}\cum{(\Delta N)}{5}
\nonumber \\ &&  {}
-15\biggl (\cum{N}{}^5+10\cum{N}{}^3\cum{(\Delta N)}{2}+10\cum{N}{}^2\cum{(\Delta N)}{3}+10\cum{(\Delta N)}{2}\cum{(\Delta N)}{3}
\nonumber \\ && \qquad \qquad  {}
+5\cum{N}{} \left(3\cum{(\Delta N)}{2}^2+\cum{(\Delta N)}{4}\right)
+\cum{(\Delta N)}{5} \biggr )
\nonumber \\ &&  {}
+\cum{(\Delta N)}{6} \,  .
\end{eqnarray}
\end{subequations}
\end{widetext}

%%%%%%%%%%%%%%%%%%%%%%%%%%%%%%%%%%%%%%%%%%%%%%%%%%%%%%%%%%%

\end{document}